\documentclass[letter]{aa} 

\newcommand{\Ni}{\ensuremath{^{56}\mathrm{Ni}}}

\newcommand{\Msun}{\ensuremath{\mathrm{M}_\odot}}
\newcommand{\Rsun}{\ensuremath{\mathrm{R}_\odot}}

\newcommand{\kmps}{\ensuremath{\mathrm{km~s^{-1}}}}

\defcitealias{mm20}{MM20}
\defcitealias{greiner2015sn11klmagnet}{G15}

\usepackage{graphicx}
\usepackage{txfonts}
\usepackage[]{hyperref}
\begin{document}

   \title{
   Luminous supernovae associated with ultra-long gamma-ray bursts from hydrogen-free progenitors extended by pulsational pair-instability
}


\authorrunning{Moriya, Marchant, and Blinnikov}
\titlerunning{Luminous SNe from extended hydrogen-free GRB progenitors}

   \author{Takashi J. Moriya
          \inst{1,2}
          \and
          Pablo Marchant\inst{3}
          \and
          Sergei I. Blinnikov\inst{4,5,6}
          }

   \institute{National Astronomical Observatory of Japan, National Institutes of Natural Sciences, 2-21-1 Osawa, Mitaka, Tokyo 181-8588, Japan\\
              \email{takashi.moriya@nao.ac.jp}
         \and
         School of Physics and Astronomy, Faculty of Science, Monash University, Clayton, Victoria 3800, Australia
         \and
             Institute of Astrophysics, KU Leuven, Celestijnenlaan 200D, 3001, Leuven, Belgium
        \and
        National Research Center "Kurchatov institute", Institute for Theoretical and Experimental Physics (ITEP), 117218 Moscow, Russia
        \and
        Space Research Institute (IKI), Russian Academy of Sciences, Profsoyuznaya 84/32, 117997 Moscow, Russia
        \and
        Kavli Institute for the Physics and Mathematics of the Universe (WPI), The University of Tokyo Institutes for Advanced Study, The University of Tokyo, 5-1-5 Kashiwanoha, Kashiwa, Chiba 277-8583, Japan        
             }

\date{Received July 13, 2020; accepted September 03, 2020}

 
\abstract{
We show that the luminous supernovae associated with ultra-long gamma-ray bursts can be related to the slow cooling from the explosions of hydrogen-free progenitors extended by pulsational pair-instability. In the accompanying paper \citep{mm20}, we have shown that some rapidly-rotating hydrogen-free gamma-ray burst progenitors that experience pulsational pair-instability can keep an extended structure caused by pulsational pair-instability until the core collapse. Such progenitors have large radii exceeding 10~\Rsun\ and they sometimes reach beyond 1000~\Rsun\ at the time of the core collapse. They are, therefore, promising progenitors of ultra-long gamma-ray bursts. We here perform the light-curve modeling of the explosions of one extended hydrogen-free progenitor with a radius of 1962~\Rsun. The progenitor mass is 50~\Msun\ and 5~\Msun\ exists in the extended envelope. We use one-dimensional radiation hydrodynamics code \texttt{STELLA} in which the explosions are initiated artificially by setting given explosion energy and \Ni\ mass. Thanks to the large progenitor radius, the ejecta experience slow cooling after the shock breakout and they become rapidly evolving ($\lesssim 10~\mathrm{days}$) luminous ($\gtrsim 10^{43}~\mathrm{erg~s^{-1}}$) supernovae in optical even without the energy input from the \Ni\ nuclear decay when the explosion energy is more than $10^{52}~\mathrm{erg}$. The \Ni\ decay energy input can affect the light curves after the optical light-curve peak and make the light-curve decay slow when the \Ni\ mass is around 1~\Msun. They also have fast photospheric velocity above 10,000~\kmps\ and hot photospheric temperature above 10,000~K at around the peak luminosity. We find that the rapid rise and luminous peak found in the optical light curve of SN~2011kl, which is associated with the ultra-long gamma-ray burst GRB~111209A, can be explained as the cooling phase of the extended progenitor. The subsequent slow light-curve decline can be related to the \Ni\ decay energy input. The ultra-long gamma-ray burst progenitors proposed in \citet{mm20} can explain both the ultra-long gamma-ray burst duration and the accompanying supernova properties. When the gamma-ray burst jet is off-axis or choked, the luminous supernovae could be observed as fast blue optical transients without accompanying gamma-ray bursts. 
}


   \maketitle
%

\section{Introduction}\label{sec:introduction}
Long-duration gamma-ray bursts (LGRBs), which are often accompanied with energetic hydrogen-free supernovae (SNe), are considered to be explosions of massive rapidly rotating Wolf-Rayet (WR) stars \citep{woosley2006rev}. Most LGRBs have a duration of the order of $1-100~\mathrm{sec}$ \citep[e.g.,][]{levan2014ulgrb}. The LGRB duration is connected with the duration of the central engine launching the jet. The central engine is suggested to be, e.g., accreting black holes (BHs, \citealt{woosley1993,paczynski1998}) or strongly magnetized neutron stars (so-called magnetars, \citealt{usov1992,duncan1992}). If the central engine is an accreting BH, the LGRB duration can be related to the accretion timescale of the collapsing WR stars that can roughly correspond to the free-fall timescale of WR stars \citep[e.g.,][]{kumar2008}. Indeed, the typical timescales of LGRBs match the free-fall timescales of WR stars that usually have a radius of the order of $1~\Rsun$ or less \citep[e.g.,][]{yoon2012popiii}.

It has been recently recognized that some LGRBs have a duration of more than $10^4~\mathrm{sec}$ and they are called ultra-long gamma-ray bursts (ULGRBs, e.g., \citealt{gendre2013bsg,levan2014ulgrb}). If the duration of more than $10^4~\mathrm{sec}$ is related to the free-fall timescale of the progenitors, they must have a radius larger than 10~\Rsun. Because the required radius is too large for WR stars, the progenitors of ULGRBs have been linked to hydrogen-rich blue supergiants (BSGs) that typically have a radius of the order of 10~\Rsun\ and have a free-fall time of more than $10^4~\mathrm{sec}$ \citep[e.g.,][]{suwa2011bsg}. However, no hydrogen signatures have ever been observed in ULGRBs \citep[][G15 hereafter]{greiner2015sn11klmagnet}. In the accompanying paper, we show that massive hydrogen-free stars that experience the pulsational pair-instability shortly before the core collapse can keep the extended structure caused by the pair-instability until the collapse and have radii well above 10~\Rsun\ (\citealt{mm20}, \citetalias{mm20} hereafter). They can even have radii exceeding 1000~\Rsun\ and sustain the accretion to the central BH for far more than $10^4~\mathrm{sec}$. The jet launched by the long-lasting accretion in the extended progenitor may also escape from the progenitor \citep{quataert2012kasen}. Thus, they are promising progenitor candidates for ULGRBs.

Not only the extremely long duration, but also the accompanying SNe in ULGRBs are distinct from those of LGRBs. SN~2011kl, which was accompanied by ULGRB~111209A, has the peak optical magnitude at around $-20~\mathrm{mag}$ and is more luminous than typical SNe associated with LGRBs (\citetalias{greiner2015sn11klmagnet}; \citealt{kann2019}). 
SN~2011kl is difficult to explain with the canonical \Ni-decay energy input for SNe associated with LGRBs because of the short rise time, high photospheric temperature, and high photospheric velocity. The luminous SNe associated with ULGRBs likely have a different luminosity source (see \citealt{ioka2016} and the references therein). For example, the magnetar powered model, which is also suggested to be the luminosity source of hydrogen-free superluminous supernovae \citep{quimby2011}, is proposed to explain the accompanying luminous SN (\citealt{metzger2015}).
SN~2011kl is also linked to the cocoon emission from a hydrogen-rich blue supergiant ULGRB progenitor \citep{nakauchi2013,kashiyama2013}, but SN~2011kl did not show the expected hydrogen signatures \citepalias{greiner2015sn11klmagnet}.

In this Letter, we investigate the observational properties of the explosions of an extended hydrogen-free ULGRB progenitor presented in \citetalias{mm20}. A systematic modeling of the explosions of the extended helium stars up to around 150~\Rsun\ has been previously performed by \citet{kleiser2018,dessart2018}, but our progenitor has a more extended envelope. We show that the luminous SN components accompanied by ULGRBs appear from our progenitor without \Ni\ or any central energy inputs thanks to the long adiabatic cooling phase caused by the large progenitor radius. The overall properties of SN~2011kl can be explained by our progenitor model. Our results show that the extended hydrogen-free ULGRB progenitors presented in \citetalias{mm20} can explain both the extremely long duration and the luminous SNe accompanied by ULGRBs, making them promising progenitors for ULGRBs.

\begin{figure}
\centering
\includegraphics[width=\linewidth]{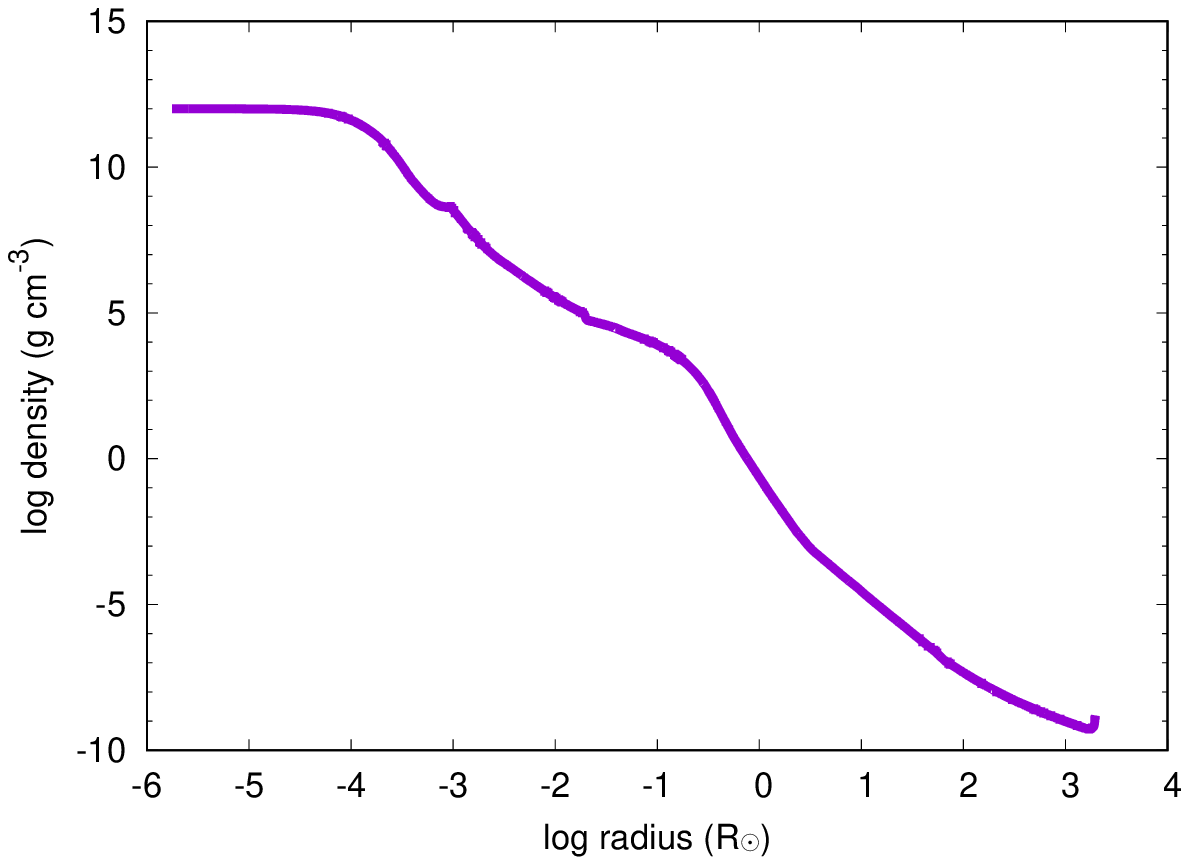}
\includegraphics[width=\linewidth]{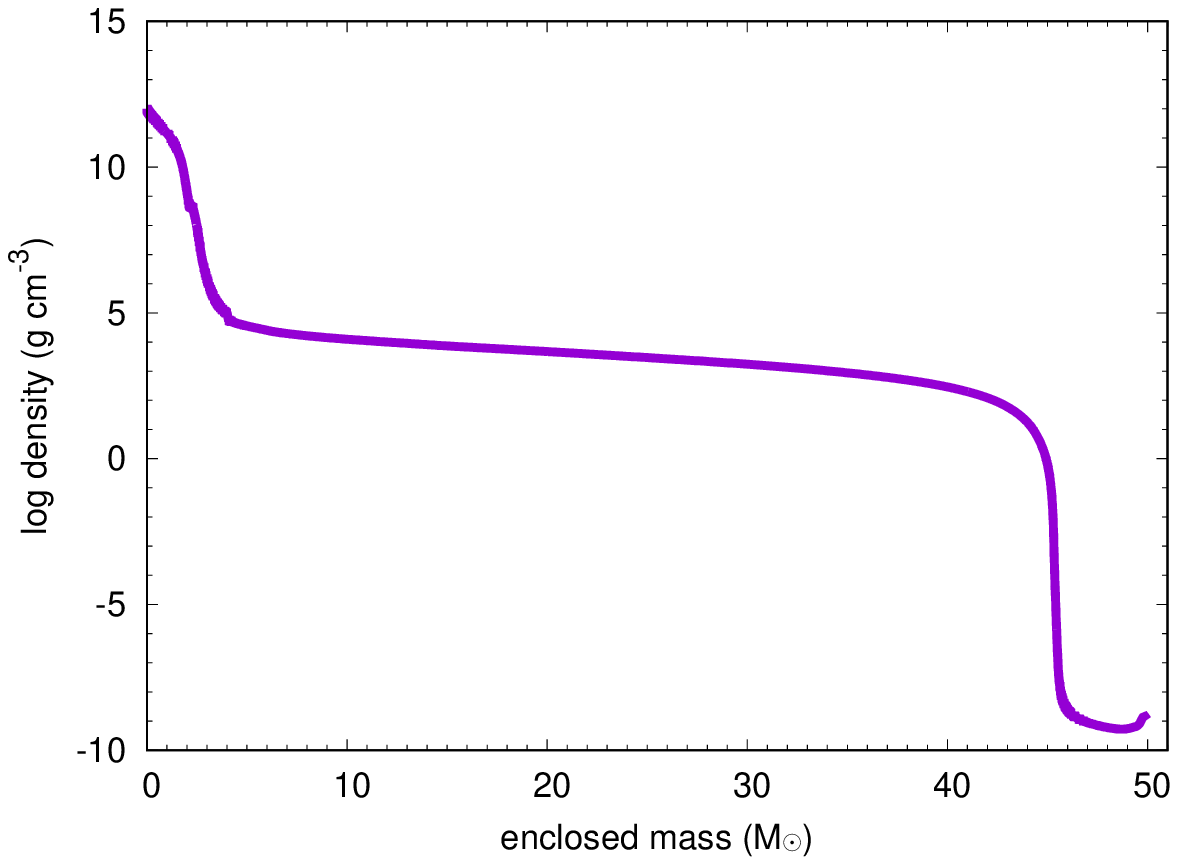}
\caption{
Density structure of the ULGRB progenitor adopted in this work. The top panel shows it in the radius coordinate and the bottom panel shows it in the mass coordinate. Note that the extended progenitor is a hydrogen-free star.
}
\label{fig:density}%
\end{figure}

\begin{figure}
\centering
\includegraphics[width=\linewidth]{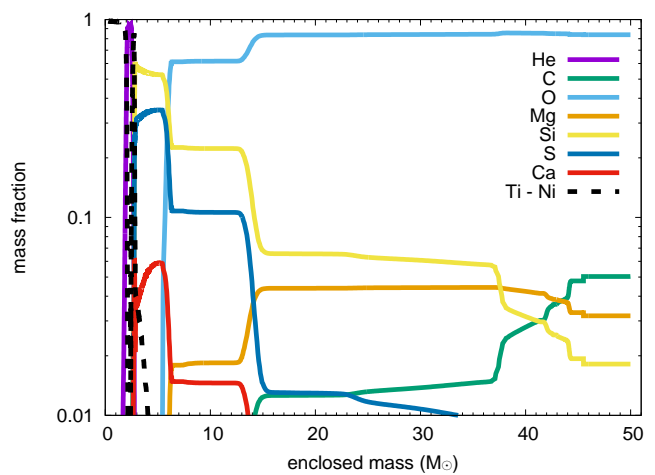}
\caption{
Abundance profile of the ULGRB progenitor.
}
\label{fig:abundance}%
\end{figure}

\section{Methods}
\subsection{Progenitor}
We take one ULGRB progenitor model from \citetalias{mm20}. We refer to \citetalias{mm20} for the details of the progenitor model. Fig.~\ref{fig:density} shows the density structure of the progenitor model. Shortly, it is initially an 82.5~\Msun\ helium star rotating with 90\% of the critical rotation. It is evolved without the Spruit-Tayler dynamo process. The progenitor experiences two pulses at 6749 and 2085 years before the final core collapse. It keeps the extended hydrostatic structure after the second pulse until the core collapse and the radius of the progenitor at the time of collapse is 1962~\Rsun. The abundance profile of the progenitor at the time of the collapse is presented in Fig.~\ref{fig:abundance}.
The progenitor, which is obtained from the one-dimensional hydrostatic stellar evolution calculation, is close to the Eddinton limit. There are significant uncertainties and undergoing works related to this regime \citep[e.g.,][]{jiang2015}. We note that the large expansion in our models does not arise from inflation in a star near the Eddington limit, but by the energy deposited in the thermonuclear pulse, which leads to a dynamical expansion by more than two orders of magnitude in a timescale where both convection and radiative transport are irrelevant \citep{marchant2019}. The exact radius of the stars could be uncertain, but the prediction that they expand significantly is reasonable.
The progenitor mass at the collapse is 50~\Msun\ and it retains enough angular momentum to be a GRB as presented in \citetalias{mm20}. The extended envelope contains 5~\Msun, which is 10\% of the progenitor mass.

The free-fall time of the progenitor is $4\times 10^7~\mathrm{sec}$ and it is well beyond the duration of ULGRBs ($\sim 10^4~\mathrm{sec}$). We presume that the accretion towards the central BH can be suppressed at some moment because of, e.g., the accompanying SN explosion or the BH accretion disk wind. The outer layers of the progenitor are less bounded and the accretion would be easily surpressed. Swift 1644+57, which is suggested to be a LGRB with the duration of $\sim 10^7~\mathrm{sec}$ \citep{quataert2012kasen}, could be the case where the BH accretion was not suppressed, for example.

\subsection{Light-curve calculations}
The numerical LC calculations are performed with the one-dimensional radiation hydrodynamics code \texttt{STELLA} \citep{blinnikov1998sn1993j,blinnikov2000sn1987a,blinnikov2006sniadeflg}. \texttt{STELLA} implicitly treats time-dependent equations of hydrodynamics and the angular moments of intensity averaged over a frequency bin with the variable Eddington method. \texttt{STELLA} calculates the spectral energy distributions (SEDs) at each time step. 

In this study, we assume an ejecta mass of 15~\Msun\ in the LC calculations and set the mass cut at 35~\Msun. This ejecta mass provided the best LC model for SN~2011kl in our investigation. The ejecta mass of around 10~\Msun\ is typically estimated in SNe associated with LGRBs \citep[e.g.,][]{iwamoto1998}. We note that the BH mass from the same progenitor is estimated as 45~\Msun\ based on its angular momentum distribution in \citetalias{mm20}, but we here set the BH mass artificially to have the selected ejecta mass. The explosions are artificially triggered by putting thermal energy required to reach the prescribed explosion energy within 0.5~\Msun\ above the mass cut. We show models with an explosion energy of 10~B, 20~B, and 50~B, where $1~\mathrm{B}\equiv10^{51}~\mathrm{erg}$. An explosion energy of 10~B and above is typically estimated in SNe associated with LGRBs \citep[e.g.,][]{woosley2006rev}. 12~\Msun\ exists below 10,000~\kmps\ in the ejecta of the 10~B explosion. We include a small amount of \Ni\ ($M_{\Ni}=0.1~\Msun$) in our standard simulations, but it does not affect the LCs we present. We also show a model with a \Ni\ mass of 1.5~\Msun. The \Ni\ mass is also set artificially.

The optical LC of SN~2011kl with which we compare our synthetic LCs is obtained by integrating the SEDs from 3500~\AA\ to 8000~\AA\ (the ``quasi-bolometric'' LC in \citetalias{greiner2015sn11klmagnet}). Therefore, we construct the optical LCs from the synthetic models by integrating the synthetic SEDs in the same wavelength range. When we show the photospheric properties, the photosphere is set at the location where the Rosseland mean optical depth becomes $2/3$.

\begin{figure}
\centering
\includegraphics[width=\linewidth]{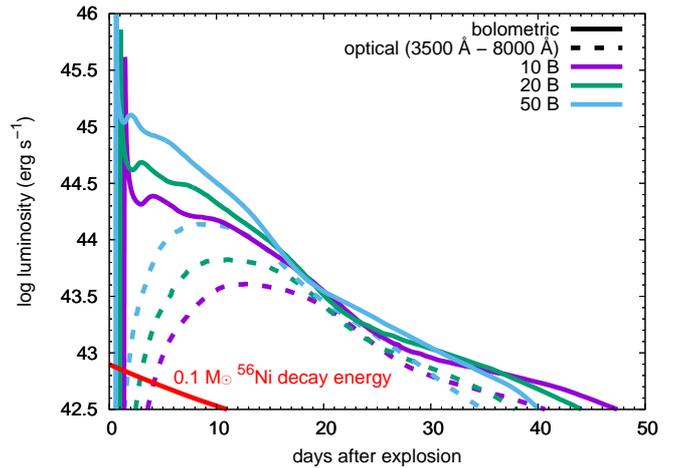}
\caption{
Bolometric (solid) and optical (dashed; integrated from 3500~\AA\ to 8000~\AA) LCs from the explosions of the extended (1962~\Rsun) hydrogen-free ULGRB progenitor. The ejecta mass is 15~\Msun\ and the explosion energy is shown in the figure. All the models have 0.1~\Msun\ of \Ni\ but it does not affect the LCs presented here as indicated by the total nuclear energy available from 0.1~\Msun\ of \Ni\ (red; \citealt{nadyozhin1994}).
}
\label{fig:bolometric}%
\end{figure}

\begin{figure}
\centering
\includegraphics[width=\linewidth]{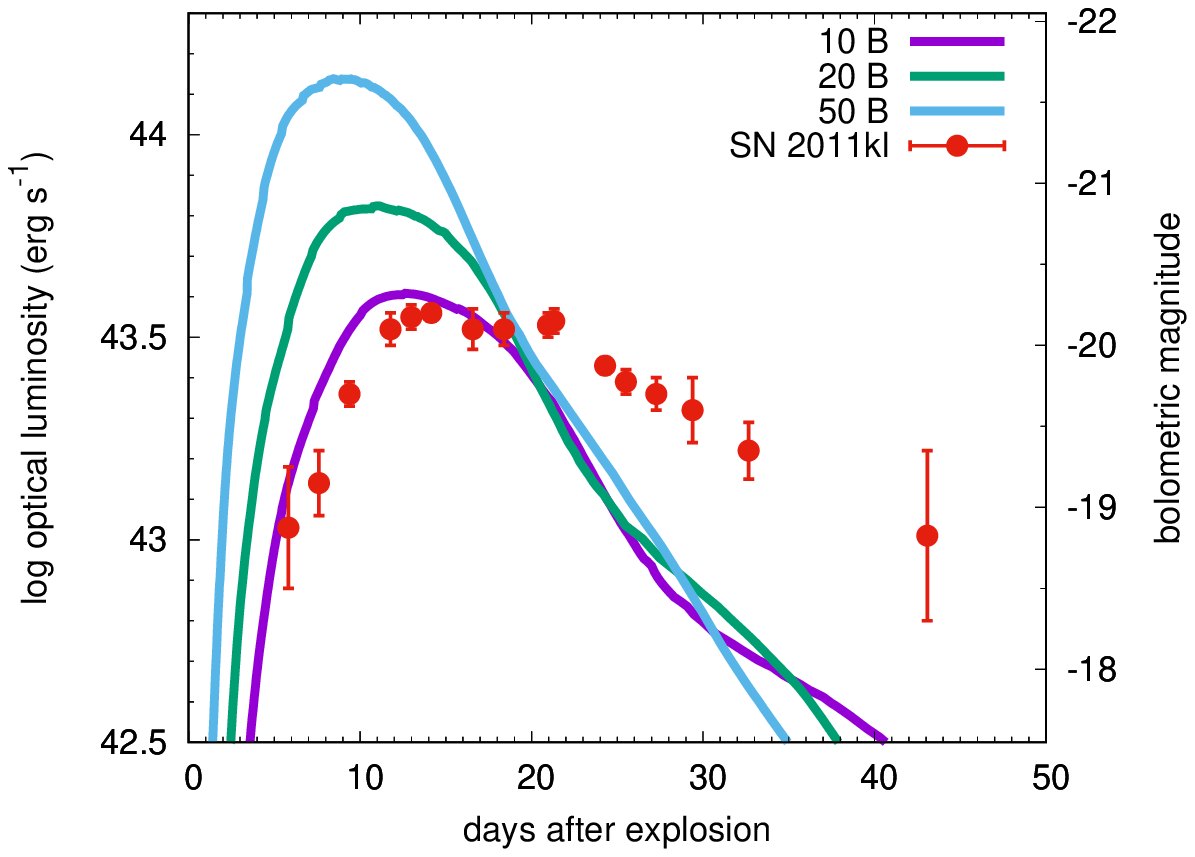}
\includegraphics[width=\linewidth]{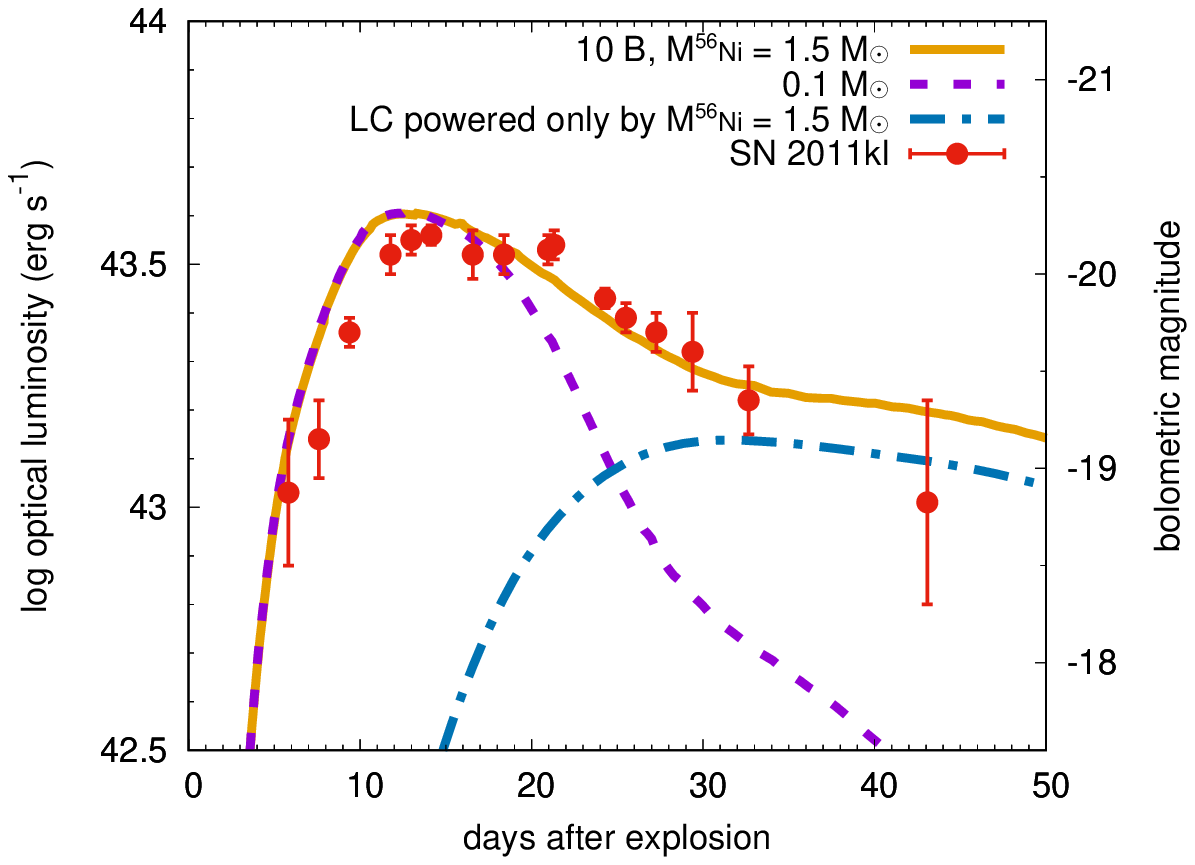}
\caption{
\textit{Top:} Synthetic optical (3500~\AA\ to 8000~\AA) LCs with $M_{\Ni}=0.1~\Msun$ compared with the optical LC of SN~2011kl. 
\textit{Bottom:}
Synthetic optical (3500~\AA\ to 8000~\AA) LC with $M_{\Ni}=1.5~\Msun$ (solid). It matches well to the observed LC of SN~2011kl. The quasi-bolometric LC with $M_{\Ni}=0.1~\Msun$ is also shown (dashed). We also present the quasi-bolometric LC model from a compact hydrogen-free progenitor having $M_{\Ni}=1.5~\Msun$, which does not have the slow cooling phase, to illustrate the luminosity contribution from the \Ni\ decay with the dot-dashed line. All the models in the bottom panel have the explosion energy of 10~B.
The explosion time for SN~2011kl is set at the time of the ULGRB~111209A trigger.
}
\label{fig:sn2011kl}%
\end{figure}

\begin{figure}
\centering
\includegraphics[width=\linewidth]{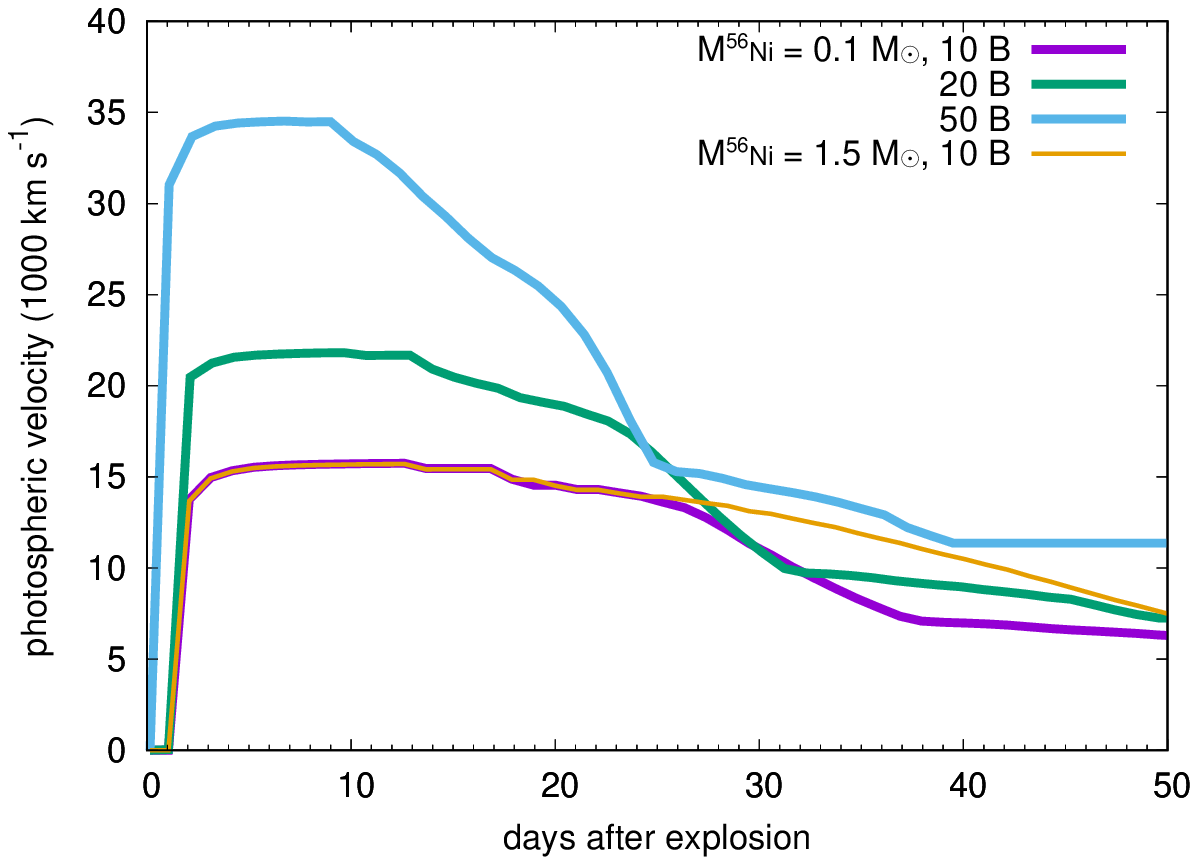}
\includegraphics[width=\linewidth]{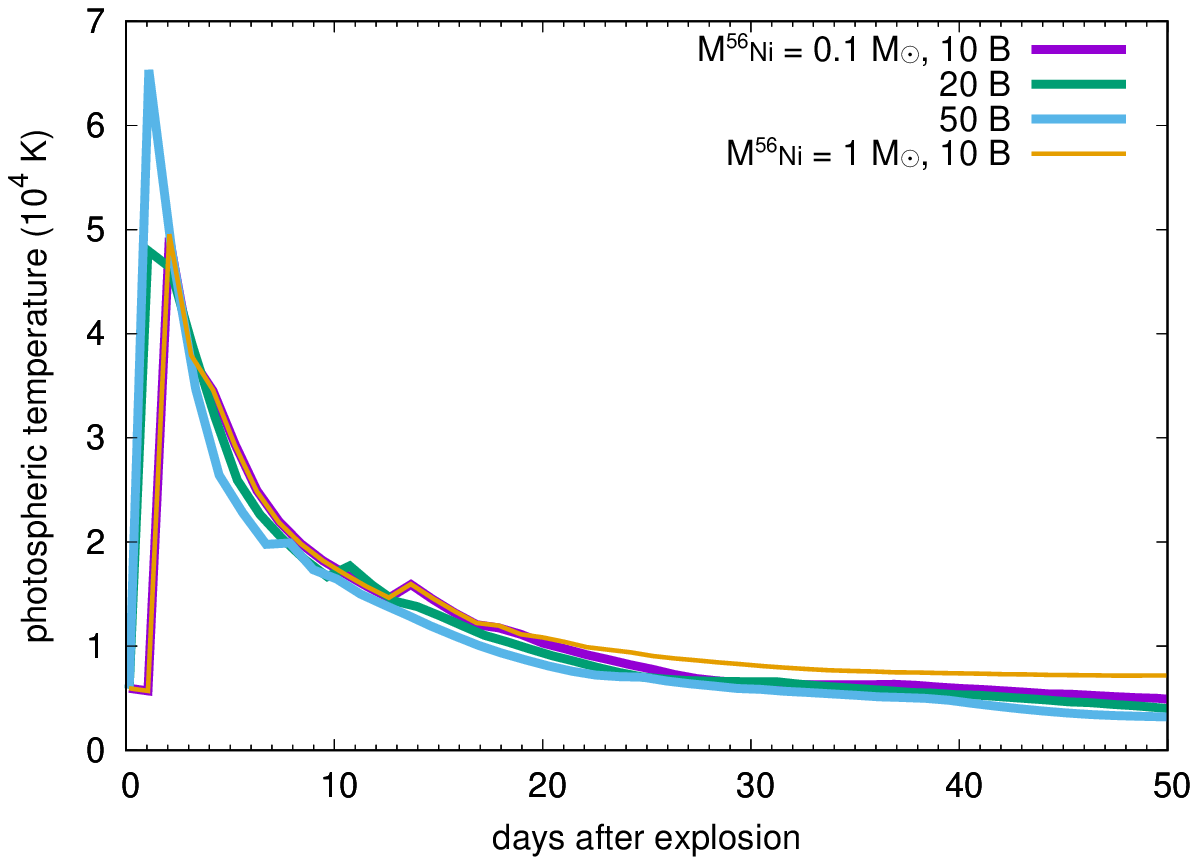}
\caption{
Photospheric velocity (top) and temperature (bottom) evolution of the models presented in this work.
}
\label{fig:photo}%
\end{figure}

\section{Results}\label{sec:results}
Fig.~\ref{fig:bolometric} shows the synthetic bolometric and optical LCs with a \Ni\ mass of 0.1~\Msun. The bolometric LC behaves in the standard way as expected in the cooling phase of the extended progenitor \citep[e.g.,][]{hoflich1993,blinnikov1998sn1993j,rabinak2011,bersten2012,nakar2014}. After the ejecta are heated by the shock wave causing the SN explosion, the ejecta experience adiabatic cooling. The bolometric luminosity steadily declines as the ejecta cool. The adiabatic cooling is less efficient in more extended progenitors and, therefore, the explosions from more extended progenitors keep a high temperature for a longer time. This means that it takes more time for the ejecta to be cool enough for optical LCs to be luminous for more extended progenitors. Thus, the optical LCs have longer rise times for larger progenitor radii. The slow cooling also makes the photosphere more extended when the optical luminosity becomes large, making more luminous optical LCs for more extended progenitors. We refer to \citet{grassberg1971,falk1977} for the further details of dynamics and LCs in the early phases of extended progenitor explosions.

The fast-evolving optical LCs here result from the cooling of the extended progenitors. The optical LCs have large peak luminosities and short rise times despite of the small \Ni\ mass and massive ejecta (15~\Msun). The 10~B explosion model, for example, reaches its optical peak luminosity of $4\times 10^{43}~\mathrm{erg~s^{-1}}$ ($-20.3~\mathrm{mag}$) in 13 days. We refer to \citet{kleiser2018,dessart2018} for the LCs from the less extended ($\lesssim 100~\Rsun$) hydrogen-free progenitors.

The optical LC of SN~2011kl, which was associated with ULGRB~111209A, is compared with our synthetic optical LCs in Fig.~\ref{fig:sn2011kl}. The 10~B explosion model well explains the early quick rise and peak luminosity of SN~2011kl but the LC decline rate after the peak is larger than that in SN~2011kl. The LC models with 0.1~\Msun\ of \Ni\ have too rapid decline after the peak. However, a larger amount of \Ni\ likely exists because of the large explosion energy. We show the LC model with $M_{\Ni}=1.5~\Msun$ in the bottom panel of Fig.~\ref{fig:sn2011kl}. The delayed luminosity input from the nuclear decay of 1.5~\Msun\ of \Ni\ successfully reproduces the slow decline in the optical LC found in SN~2011kl. \citetalias{greiner2015sn11klmagnet} disfavored the possibility that SN~2011kl is powered by the \Ni\ decay because of the short rise. We here show that the quick rise time is reproduced by the extended envelope and then the \Ni\ decay can contribute to the declining phase. The existence of the extended envelope naturally explains the ultra-long duration of GRB~111209A at the same time.

Fig.~\ref{fig:photo} presents the photospheric velocity and temperature evolution of our explosion models. We can find that our luminous SN models are characterized by large photospheric velocities of the order of 10,000~\kmps\ and hot photospheric temperature exceeding $10,000~\mathrm{K}$ at around the LC peak. The photospheric velocity of SN~2011kl at around the peak luminosity is estimated to be 19,000~\kmps\ based on the calcium and carbon lines \citepalias{greiner2015sn11klmagnet}. Our best LC model for SN~2011kl presented in the bottom panel of Fig.~\ref{fig:sn2011kl} has the photospheric velocity of 15,000~\kmps\ at around the LC peak. We estimate the photosphere based on the Rosseland mean opacity. It does not necessarily correspond to where calcium and carbon is absorbed and they could be absorbed above the photosphere. In addition, the photosphere in the model at around the LC peak locates near the outer edge of the ejecta and the photospheric velocity would be sensitive to the structure of the outer layers as well. We, therefore, argue that the photospheric velocity we obtain in the model is consistent with the observation of SN~2011kl. In any case, our model is shown to have very large photospheric velocities close to the observed one. Overall, our extended hydrogen-free progenitor model explains the observed properties of SN~2011kl well.

\section{Discussion}\label{sec:discussion}
We show that the hydrogen-free GRB progenitor model extended by the pusational pair-instability can be accompanied by luminous fast-evolving SNe in optical. Our 10~B explosion model has the peak optical luminosity of $4\times 10^{43}~\mathrm{erg~s^{-1}}$ and the rise time of 13~days in optical. The optical LC properties match well to those of SN~2011kl. When ULGRBs are triggered by progenitors of the order of 1000~\Rsun, we are likely to observe the luminous SN component. The radius extension by the pulsational pair-instability can also result in less extended ULGRB progenitors of the order of 100~\Rsun\ or less \citepalias{mm20}. Such less extended progenitors are expected to make less luminous ($\lesssim 10^{43}~\mathrm{erg~s^{-1}}$) faster evolving ($\lesssim 1~\mathrm{days}$) SNe in optical. However, such early SN components are likely fainter than the GRB afterglow components and may be difficult to identify. ULGRBs without the SN component may, therefore, originate from extended progenitors with a radius of the order of 100~\Rsun\ or less. The difference in the progenitor radii can explain why the luminous SN components in optical appear only in some ULGRBs.

The luminous SNe we presented here are not necessarily accompanied with ULGRBs because ULGRBs can be off-axis. Alternatively, the jet launched in the progenitors could be choked and the explosions of some ULGRB progenitors may occur without GRBs \citep[e.g.,][]{milisavljevic2015choked}. The quickly-rising, luminous, blue SNe we predict here match the properties of fast blue optical transients (FBOTs, \citealt{drout2014fast,tampo2020}). They have a rise time of less than 10~days and peak luminosities ranging from $\sim 10^{42}~\mathrm{erg~s^{-1}}$ to $\sim 10^{44}~\mathrm{erg~s^{-1}}$. These properties are consistent with our LC models \citep[cf.][]{kleiser2014,kleiser2018}. Our LC models are also consistent with those of rapidly rising luminous transients having peak luminosities between $\simeq -20$~mag and $\simeq -21$~mag \citep{arcavi2016rapidrise}. Some of these transients can also be related to the explosions of the hydrogen-free stars extended by pulsational pair-instability.

The FBOTs and luminous SNe, including superluminous SNe, are often related to the birth of magnetars \citep[e.g.,][]{metzger2015}. Our model presented here does not exclude the magnetar scenario to explain these transients. For SN~2011kl, the quick rise and luminous peak in optical can be explained by our extended progenitor explosion but the additional energy source is required to explain the subsequent slow LC decline. The magnetar model can explain the optical LC without additional energy source, but it is argued that powering both GRB and SN at the same time with a single magnetar is difficult \citep{ioka2016}. We here show that the fast-evolving luminous SNe associated with ULGRBs can be explained even if ULGRBs are triggered with the BH accretion. In our model, the optical luminosity increase is caused by the cooling and the ejecta are much hotter before the optical luminosity increase, while the magnetar powered models are likely predict a cooler temperature before the optical luminosity increase caused by the heating from the magnetar spin down near the center of the ejecta. The two models could be distinguished by the early ultraviolet observations that can constrain the temperature before the optical LC peak.

In this work, we do not consider the effect of circumstellar matter (CSM) on the LCs. Pulsational pair-instability leads to mass ejection from the progenitors. The extended cool envelope may also experience enhancement in mass loss as found in red supergiants \citep[cf.][]{vanloon2005}. The luminous LC component caused by the extended progenitor can be followed by the interaction-powered LC component. For example, we suggest that the slow decline of SN~2011kl can be due to the \Ni\ decay in the previous section. Such a slow decay may also be realized by the interaction between the ejecta and the dense CSM \citep[cf.][]{moriya2018slsn}.

The progenitor model we adopted in this work experiences the last pulse 2085~years before the core collapse. In other cases, however, the final pulse can occur decades or years before the core collapse \citepalias{mm20}. If the pulsational pair-instability explosions shortly before the core collapse leading to ULGRBs or FBOTs become bright enough \citep[e.g.,][]{woosley2007ppisn}, they can be observed as precursor transients of ULGRBs or FBOTs.

Finally, if we are lucky enough to have a direct progenitor detection of a ULGRB progenitor, it will be observed as a blue or red supergiant star that does not have hydrogen and is dominated by carbon and oxygen. For example, the surface effective temperature of the hydrogen-free progenitor we investigated in this work is only 5228~K.

\section{Conclusions}\label{sec:conclusions}
We showed that the explosions of the rapidly rotating hydrogen-free GRB progenitor extended by pulsational pair-instability result in luminous, rapidly rising, blue, fast SNe in optical. The hydrogen-free progenitor investigated in this work has 1962~\Rsun\ at the time of the collapse. The slow cooling of the extended progenitor makes such luminous SNe. The rise time is typically $\lesssim 10~\mathrm{days}$ and peak luminosity can be more luminous than $-20~\mathrm{mag}$, depending on the explosion energy. They have the photospheric velocity well above 10,000~\kmps\ and the photospheric temperature exceeds 10,000~K at around the LC peak. We find that the quick rise and large peak luminosity of the luminous SN~2011kl associated with ULGRB~111209A can be well explained by the explosion of our extended progenitor. Our LC model declines faster than SN~2011kl but the late-phase LCs could be additionally powered by the \Ni\ decay or the CSM interaction. When the GRB jet is off-axis or chocked, such luminous SNe can be observed without accompanying GRBs. Some of FBOTs and other rapidly rising transients may correspond to such cases. The final pulse due to the pulsational pair-instability can occur $\sim 1 - 1000$~years before the core collapse. If the final pulse occurs decades or years before the core collapse and it gets bright enough, it is possible that we detect a precursor transient before an ULGRB or FBOT.

\begin{acknowledgements}
T.J.M. is supported by the Grants-in-Aid for Scientific Research of the Japan Society for the Promotion of Science (JP18K13585, JP20H00174).
P.M. acknowledges support from the FWO junior postdoctoral fellowship No. 12ZY520N.
S.B. has been supported by the grant RSF 18-12-00522 in
his work on the supernova and GRB-AG simulations.
Numerical computations were in part carried out on PC cluster at Center for Computational Astrophysics (CfCA), National Astronomical Observatory of Japan.
\end{acknowledgements}

%
   \bibliographystyle{aa} 
   \bibliography{aanda.bib} 
%

\end{document}